\documentstyle[12pt]{article}
\raggedright 
\renewcommand{\section}[1]{\refstepcounter{section}
\vspace{24pt}\noindent{\bf\arabic{section}.\quad #1}
\vspace*{12pt}}
\newcommand{\ulsect}[1]{\vspace{18pt}\noindent{\bf #1}
\vspace*{12pt}}

\begin{document}
\begin{flushright} KSUCNR-011-92\\
TPI-MINN-92/47-T\\
(revised)\\
\end{flushright}
\vspace*{5mm}
\begin{center}
{\bf Correlation measurements in high--multiplicity events}\\[10mm]
David Seibert \\[5mm] Physics Department$^*$ \\ Kent State
University, Kent, OH 44242 \\[3mm] and \\[3mm] Theoretical Physics
Institute \\ University of Minnesota, Minneapolis, MN 55455\\[10mm]
{\bf Abstract} \\
\end{center}
Requirements for correlation measurements in high--multiplicity events
are discussed.  Attention is focussed on detection of
so--called hot spots, two--particle rapidity correlations,
two--particle momentum correlations (for quantum interferometry) and
higher--order correlations.  The signal--to--noise ratio may
become large in the high--multiplicity limit, allowing meaningful
single--event measurements, only if the correlations are due to
collective behavior.
\vfill
\begin{center}
{\em to be published in Phys.\ Rev.\ C.}
\end{center}
\vspace*{10mm}
\begin{flushleft} KSUCNR-011-92\\
TPI-MINN-92/47-T\\
October 1992\\
(revised January 1993)\\ \end{flushleft}
\thispagestyle{empty}\mbox{}
\footnoterule
{$^*$On leave until October 12, 1993 at: Theory Division, CERN,
CH--1211 Geneva 23, Switzerland; Internet: seibert@surya11.cern.ch.}
\newpage
\setcounter{page}{1} \pagestyle{plain}

\section{Introduction} \label{sintro}

In the next ten years, ultra--relativistic heavy ion collisions are
planned at Brookhaven's Relativistic Heavy Ion Collider (RHIC) and CERN's
Large Hadronic Collider (LHC).  In both cases, experimenters expect to see
multiplicities in excess of 1000 particles per unit rapidity.  As a
result, there has been a lot of speculation about single--event
fluctuation measurements [\ref{rsea}--\ref{rseo}].  In this paper, I
assess requirements for various single--event measurements, and calculate
the number of events needed for useful measurements in cases where
single--event analyses are not meaningful.

I begin with a discussion of searches for so--called hot
spots (regions of unusually high particle density) in
Section~\ref{slfhs}.  In Section~\ref{stprc}, I discuss measurements
of two--particle rapidity correlation functions, assuming that
high--multiplicity events are independent superpositions of
lower--multiplicity events.  In Section~\ref{sqi}, I discuss measurements
of two--particle momentum correlation functions that are commonly
constructed to measure collision volumes using quantum interferometry.
In Section~\ref{shocf}, I compare results from two--particle correlation
functions with those from higher--order correlation functions.  Finally,
I summarize the results in Section~\ref{sc}.

\section{ Looking for hot spots } \label{slfhs}

One common suggestion is that it may be possible to search for
hot spots, or regions with unusually large numbers of pions.  The
basic motivation for these searches is simple: any process that creates
a lot of entropy in a small rapidity bin is of interest.
Thus, hot spots are commonly thought of as possible signals for
interesting phenomena.

For definiteness, suppose that searches are made within a rapidity
window of size $\Delta Y$, using events with $N$ particles in this
window.  If particles are randomly distributed with
a flat rapidity distribution, the mean number of particles in a bin
of size $\delta y$ is
\begin{equation}
\overline{n} ~=~ \frac {N \, \delta y} {\Delta Y} ~=~ (dN/dy) \,
\delta y,
\end{equation}
where $dN/dy = N/\Delta Y$.  The standard deviation is
\begin{equation}
\sigma_n ~=~ \overline{n^2} \, - \, \overline{n}^2 ~=~ \overline{n} \,
(1-\delta y/ \Delta Y) ~\approx~ \overline{n},
\end{equation}
if the window is large ($\Delta Y \gg \delta y$).

The central limit theorem applies in the limit $N \rightarrow \infty$,
so particle number fluctuations have a gaussian distribution.  The
probability that a given bin contains more than $\overline{n} +
\delta n$ particles is
\begin{equation}
{\cal P} (\delta n / \overline{n}^{1/2}) ~\approx~ \left( \frac
{\overline{n}} {2\pi \, \delta n^2} \right)^{1/2} \, e^{-\delta n^2 / 2
\overline{n}},
\end{equation}
if $\delta n^2 \gg \overline{n}$.  To achieve success in a hot spot
search, the hot spots must be present significantly more often than in a
random distribution:
\begin{equation}
f(\delta n) ~\gg~ {\cal P} (\delta n/\overline{n}^{1/2}),
\end{equation}
where $f(\delta n)$ is the probability that a given bin contains a hot
spot with at least $\delta n$ excess particles.

For definiteness, I assume that a useful result must find hot spots at
least ten times as often as expected.  In this case, hot spots that occur
in 10\% of the bins must produce at least $2.4\overline{n}^{1/2}$ excess
particles.  For RHIC and LHC events, particles are produced over about
ten units of rapidity, so hot spots that occur in more than 10\% of bins
will be seen more often than once per event, and are thus not very useful
as triggers for interesting events.
The larger $dN/dy$ is, the larger the hot spots must be before they can
be separated from the background fluctuations, so it is likely that hot
spot searches will be most profitable in events of high energy but
relatively low multiplicity.

For RHIC and LHC experiments, rapidity densities in excess of $dN/dy =
1000$ are expected.  A hot spot that produces high energy pions ($p > m$
in the hot spot rest frame) isotropically has a width of at least one
unit of rapidity, so I take $\delta y = 1$.  In this case, if hot spots
occur in 10\% of the bins (about one hot spot per event), they must
produce 75 excess charged pions to be clearly useful as a trigger.  If
they occur in 1\% of the bins (about one per 10 events), they must produce
100 excess pions.

One possible mechanism for visible hot spots is the production of bubbles
of disordered chiral condensate [\ref{rkt}].  These bubbles were proposed
as an explanation for the so--called Centauro events observed in
cosmic ray studies, in which many charged pions were produced with
very few neutral particles, in contrast to typical nuclear and
high--energy processes that produce equal numbers of $\pi^+$, $\pi^-$,
and $\pi^0$ mesons.  A chirally--disordered bubble that produces $N^{\pm}$
charged pions, with no $\pi^0$ mesons, yields $N^{\pm}/3$ excess charged
pions within one unit of rapidity of the bubble.  If these bubbles are
produced in less than 10\% of the bins, they will be clearly visible
if $N^{\pm} > 225$, but this is unlikely as the expected value is
$N^{\pm} \approx 20$ [\ref{rkt}].  Thus, hot spot searches at RHIC and
LHC probably will not yield evidence for bubbles of disordered chiral
condensate.

Quark--gluon plasma (QGP) droplets are typically too small to be hot spot
candidates in high--multiplicity events.  A typical mean radius for a
QGP droplet is 1 fm, at which size the droplet should produce about 18
charged pions.  There is no proposed mechanism that would produce hot
spots large enough to be clearly observable at RHIC or LHC; however, if
they are seen, the lack of theoretical prediction will make them even
more interesting than if they were expected.

\section{ Two--particle rapidity correlations } \label{stprc}

In this section, I discuss measurements of two rapidity correlation
functions: the standard two--particle correlation function, $R_2$, and
the simplest split--bin correlation function, $S_2$.
The standard two--particle rapidity correlation function is [\ref{raa},
\ref{rr2}]
\begin{equation}
R_2 (y; \Delta Y) = \frac {\rho^{(2)}(0, y)} {\rho^{(2)}(0,\Delta Y)},
\label{eR2}
\end{equation}
where $\Delta Y$ is some large rapidity separation that is used as a
reference.  I assume at first that $\rho$ is flat, and discuss the
effect of corrections for non--flat distributions afterwards.

Suppose that I take some arbitrary model of particle production and
analyze a single event.  Consider events with $N$ particles in a
rapidity window of width $\Delta Y$, where the rapidity distribution
for any given particle is $p(y) \, = \, 1/\Delta Y$.  Let there be $N_c$
correlated pairs, where typically $N_c \ll N(N-1)$, and the rapidity
distribution for correlated pairs is $q(y_1-y_2)/\Delta Y$, where $q$ is
some arbitrary function.  For simplicity, I consider only events with
exactly $N$ particles in the rapidity window; generalization to events
with differing multiplicities is straightforward [\ref{rmult}].

I do not assume that all correlations are pair--wise, but I neglect
higher--order correlations for the moment.  For example, it is possible
that the pair--wise correlations result from interactions of large
numbers of particles.  Even in this case, however, correlation functions
are dominated by pair--wise correlations unless the interactions involve
almost all of the particles.  I discuss this in more detail in
Section~\ref{shocf}.

Finally, I assume that a superposition of $a$ independent events with
$n$ particles each is equivalent to a single event with $an$ particles.
This is equivalent to assuming independent nucleon collisions, or
independent parton collisions.  In this case I must have $N_c~=~kN$,
where $k$ is some unknown proportionality constant.  [For example, if I
combine two events I double both $N$ and $N_c$, as pairs of particles
from different events are clearly uncorrelated.]

I can immediately write down the two--particle density,
\begin{equation}
\rho^{(2)} (y_1, y_2) = \frac {[N(N-1) -N_c] ~+~ N_c \, \Delta Y \,
q(y_1-y_2)} {\Delta Y^2}. \label{erho2}
\end{equation}
Using Eq.~(\ref{erho2}), I obtain the two--particle correlation function,
\begin{equation}
R_2(y; \Delta Y) ~=~ 1 \, + \, \frac {k} {dN/dy} \, \left[ q(y)
- q(\Delta Y) \right]. \label{e6}
\end{equation}
Here (and for the remainder of this paper) I drop corrections of order
$1/N$ and $1/\Delta Y$ unless otherwise specified, as I am primarily
interested in the analysis of high--multiplicity, high--energy events.
If the mean separation for a correlated pair is $y^*$, then typically
$q(0) \approx 1/y^*$, while $q(\Delta Y) \rightarrow 0$ for large
$\Delta Y$, so the maximum value of $R_2$ is roughly
\begin{equation}
R_2^{\rm max} ~\approx~ 1 \, + \, \frac {k} {(dN/dy) y^*}. \label{e7}
\end{equation}

I obtain a lower limit for the error in a measurement of $R_2$ by
calculating the expected fluctuations in the absence of correlations.
Consider an experimental measurement:
\begin{equation}
R_2(y; \Delta Y) = \frac {\displaystyle \sum_{i=1}^{N_{ev}} \, n_i(0) \,
n_i(y)} {\displaystyle \sum_{i=1}^{N_{ev}} \, n_i(0) \, n_i(\Delta Y)}.
\end{equation}
Here $n_i(z)$ is the number of particles in the $i$--th event with
rapidities between $z-\delta y/2$ and $z+\delta y/2$ ($\delta y$ is thus
the experimental bin size), and $N_{ev}$ is the number of events used in
the measurement.

Assuming that $\delta y \leq y$, so that the bins do not overlap, and
that there are no correlations,
\begin{equation}
\langle n_i(0) \, n_i(y) \rangle = \frac {N (N-1) \, \delta y^2}
{\Delta Y^2},
\end{equation}
independent of $y$, so clearly $\langle R_2 \rangle = 1$.  The standard
deviation is
\begin{equation}
\sigma_R \, = \, \frac {\displaystyle \sum_{i=1}^{N_{ev}}
\, n_i^2(0) \, n_i^2(y)} {\displaystyle \left[ \sum_{i=1}^{N_{ev}} \,
n_i(0) \, n_i(y) \right]^2} \, + \, \frac {\displaystyle \sum_{i=1}^{N_{ev}}
\, n_i^2(0) \, n_i^2(\Delta Y)} {\displaystyle \left[ \sum_{i=1}^{N_{ev}} \,
n_i(0) \, n_i(\Delta Y) \right]^2} \, - \, \frac {2} {N_{ev}}.
\end{equation}
In the absence of correlations, the standard deviation is
\begin{equation}
\sigma_R \, = \, \frac {(-8N+12) \delta y^2 + 4(N-2)
\Delta Y \delta y + 2 \Delta Y^2} {N_{ev} N(N-1) \delta y^2}.
\end{equation}

For a good measurement, the experimental bin size must be much smaller
than the total rapidity window used, so $\delta y \ll \Delta Y$.  As the
purpose of this paper is to consider measurements in high--multiplicity
events, I take the limit $N \rightarrow \infty$, and obtain
\begin{equation}
\sigma_R \, = \, \frac {4 \, \Delta Y} {N_{ev} \, N \, \delta y}
\, \left[ 1 \, + \, {\cal O} \left( \frac {\delta y} {\Delta Y} \right)
\right].
\end{equation}
Thus, the error in the measurement is
\begin{equation}
e_R \, = \, 2 / \sqrt{N_{ev} \, (dN/dy) \, \delta y},
\label{eeR2} \end{equation}
where $dN/dy = N/\Delta Y$ is the rapidity density.

It is possible, if the fluctuations in the system are large, that the
actual error is larger than given by (\ref{eeR2}).  If the measured
standard deviation is smaller, however, then the value of
eq.~(\ref{eeR2}) is probably better to use, as this represents the error
in measuring uncorrelated events.  If the measured fluctuations are
anomalously small, then the system is probably strongly correlated, so
if $R_2-1$ is not significantly different from zero it is probably best
to look for another correlation function that reflects these strong
correlations.

Combining eqs.~(\ref{e7}) and (\ref{eeR2}), I obtain the
signal--to--noise ratio,
\begin{equation}
(s/n)_R \, = \, \frac {k} {2 y^*} \, \left( \frac {N_{ev} \, \delta y}
{dN/dy} \right)^{1/2}.
\end{equation}
If all particles are produced in clusters containing $n_c \gg 1$
particles, and all particles in a given cluster are pair--wise correlated,
then $k \approx n_c$.  [This is trivial -- every particle is produced with
$n_c$ associated (correlated) particles, so there are $n_c$ correlated pairs
per particle.]  Finally, for a passable measurement $\delta y <
y^*$, so the best possible signal--to--noise ratio is
\begin{equation}
(s/n)_R \, = \, \frac {n_c} {2} \, \left( \frac {N_{ev}} {(dN/dy) \,
y^*} \right)^{1/2}. \label{esn1}
\end{equation}

I use eq.~(\ref{esn1}) to estimate how many events I need in order to
measure $R_2$ well, as a function of the cluster size.  For most clusters,
$y^* < 1.3$, so to measure $R_2$ (at its peak) with better than $4\sigma$
accuracy I need
\begin{equation}
\frac {n_c} {2} \, \left( \frac {N_{ev}} {1.3 \, dN/dy} \right)^{1/2} > 4,
\end{equation}
or
\begin{equation}
N_{ev} \, > \, 83 \, (dN/dy) \, / \, n_c^2.
\end{equation}
Here I use $y^*=1.3$; this is obtained for the most energetic clusters,
and gives the most pessimistic estimates of $s/n$.
The cluster size seen in nuclear collisions [\ref{rklm}, \ref{remu01}]
at 200 GeV is approximately ten charged particles [\ref{rcl1}].  If this
persists up to RHIC and LHC collision energies, where the charged
particle multiplicity $dN_{ch}/dy \approx 1000$, then approximately 830
events will be needed to obtain a good measurement of the peak value of
$R_2$.

This is, of course, a naive theorist's estimate, leaving out any
possible experimental difficulties, and applies only to a measurement of
the amplitude of $R_2$.  If I want to measure the shape of $R_2$
reasonably well, I should really require that $\delta y < y^*/10$, in
which case I find that I need 8,300 events.  However, eq.~(\ref{esn1})
does illuminate the difficulty of measuring correlation functions in
high--multiplicity events: if cluster (or source) sizes are independent
of $dN/dy$, then $s/n$ decreases with increasing multiplicity, and
accurate measurement becomes increasingly difficult.  Also,
eq.~(\ref{esn1}) justifies {\it a posteriori} the neglect of correlations
when calculating the measurement error, as for one event the correlations
(the signal) are much smaller than the statistical fluctuations (the
noise).

One could argue that the measurement I have outlined for $R_2$ does not
efficiently use the available statistics.  As a response, I construct
split--bin correlation functions (SBCFs) [\ref{rsbcfs}], in order to use
the available statistics with maximal efficiency.  The simplest
second--order SBCF is
\begin{equation}
S_2(\delta y; \Delta Y) = \frac {\displaystyle \Delta Y \,
\sum_{j=1}^{\Delta Y/\delta y} \, \int_{(j-1) \delta y}^{(j-1/2) \delta y}
dy_1 \, \int_{(j-1/2) \delta y}^{j \delta y} dy_2 \, \rho^{(2)} (y_1, y_2)}
{\displaystyle \delta y \, \int_{0}^{\Delta Y/2} dy_1 \,
\int_{\Delta Y/2}^{\Delta Y} dy_2 \, \rho^{(2)} (y_1, y_2)}.
\end{equation}
Here $\Delta Y/\delta y$ must be an integer; taking $\delta y=\Delta Y/2^i$
for $i=0,1,2,\ldots$ uses all of the two--particle phase space without
re--using any pairs of particles.

Under the assumptions that I used to calculate $R_2$,
\begin{equation}
S_2 (\delta y; \Delta Y) \, = \, 1 \, + \, \frac {k} {dN/dy} \,
\left[ g(\delta y) - g(\Delta Y) \right], \label{eS20}
\end{equation}
where
\begin{equation}
g(z) \, = \, \frac {4} {z^2} \, \left[ \int_0^{z/2} dx \, x \, q(x)
\, + \int_{z/2}^z dx \, (z-x) \, q(x) \right].
\end{equation}
For $\delta y < y^*$, if $q(z)$ is linear in $z$,
\begin{equation}
g (\delta y) ~=~ q (\delta y/2), \label{es2l}
\end{equation}
while for quadratic $q(z)$,
\begin{equation}
g (\delta y) ~=~ q (\delta y/\sqrt{24/7}). \label{es2q}
\end{equation}
Eq.~(\ref{es2q}) is very close to eq.~(\ref{es2l}), so these relations
are insensitive to the shape of $q$ and are thus robust. For $\Delta Y
\gg 2y^*$,
\begin{equation}
g (\Delta Y) ~=~ \frac {4 y^*} {\Delta Y^2} ~\rightarrow~ 0, \label{egl0}
\end{equation}
so for $\Delta Y \gg 2y^*$ and $\delta y < y^*$,
\begin{equation}
S_2 (\delta y; \Delta Y) ~\approx~ R_2 (\delta y/2),
\end{equation}
independent of $\Delta Y$.  The maximum value of $S_2 (\Delta Y)$ is then
\begin{equation}
S_2^{\rm max} (\Delta Y) \, = \, 1 \, + \, \left[ 1 - \frac {4 {y^*}^2}
{\Delta Y^2} \right] \, \frac {k} {(dN/dy) \, y^*}. \label{eglarge}
\end{equation}

I calculate the error in the same manner as before, assuming that there
are no correlations:
\begin{equation}
\sigma_S \, = \, \frac {4} {N_{ev} \, N} \, \left[ 1 \,
+ \, {\cal O} \left( \frac {\delta y} {\Delta Y} \right) \right].
\end{equation}
Thus, the error in the measurement is
\begin{equation}
e_S \, = \, 2 / \sqrt{N_{ev} \, (dN/dy) \, \Delta Y},
\label{eeS2} \end{equation}
and the signal--to--noise ratio is
\begin{equation}
(s/n)_S \, = \, \frac {k} {2 \, y^*} \, \left( \frac {N_{ev} \, \Delta Y}
{dN/dy} \right)^{1/2}, \label{esonS}
\end{equation}
for $\Delta Y \gg 2 y^*$.  Setting $k=10$, $y^*=1.3$, $\Delta Y=10$ and
$dN/dy=1000$ as previously, I need approximately 110 events to measure
$S_2$ to 4$\sigma$ accuracy, as compared with 830 events to determine
the peak value of $R_2$.

The reader should note that this is an estimate of the {\em minimum}
requirements for a measurement of $S_2$.  Applying eq.~(\ref{esonS}) to
a sample of 92 central O+Em events at 200 GeV, with $y^*=1.3$, $dN/dy
\approx 40$, and $\Delta Y=4$, I obtain $(s/n)_S \approx 12$.  If
I use eq.~(\ref{eglarge}) to estimate the corrections for the finite
value of $\Delta Y$, I obtain $(s/n)_S \approx 7$.  This is in reasonable
agreement with the observed value $(s/n)_S \approx 2-3$ [\ref{rs2d}].
Most of the difference comes from the crude approximation used for
$R_2^{\rm max}$, as an exact calculation [\ref{rRS}] shows that
eq.~(\ref{e7}) overestimates the signal by a factor of two.
Thus, the estimate of $s/n$ is within a factor of 3, while the number of
events needed for a good measurement is a factor of about 10 more than
estimated.

The measurement of $S_2$ discussed above would give the shape of $R_2$
with points that are approximately 0.7 apart on a logarithmic scale, as
I keep changing the bin size by factors of two.  The previously discussed
measurement of $R_2$ gave points approximately 0.1 apart (spacing was
$y^*/10$).  Duplication of this measurement using $S_2$ would involve
seven independent measurements of $S_2$, requiring approximately 800
events, as opposed to the 8,300 required for the $R_2$ measurement.
Thus, if the same number of events are used in each case, a measurement
of $S_2$ has a bit less than one--third of the statistical noise of the
corresponding measurement of $R_2$.

If the single--particle distribution is not flat, the correlation
functions, $R_2$ and $S_2$, should be modified.  The most useful
two--particle correlation function is
\begin{eqnarray}
R_2' (y; \Delta Y) = \frac {\rho^{(2)}(0, y) \, \rho(\Delta Y)}
{\rho^{(2)}(0,\Delta Y) \, \rho(y)}, \label{eR2p} \\
\simeq \frac {\rho^{(2)}(0,y) \, \langle N \rangle^2}
{\rho(0) \, \rho(y) \, \langle N(N-1) \rangle}. \label{eR2pp}
\end{eqnarray}
These rapidity correlation functions are unity whenever particles
are distributed randomly in rapidity according to the single-particle
distribution, independent of the multiplicity distribution.
I outline the revised calculations for $R_2$ below.

Suppose that the distribution of cluster centers is $\pi \left(
(y_1+y_2)/2 \right)$, instead of simply $1/\Delta Y$ as assumed so far.
If all particles come from clusters, then the shape of the
single--particle distribution is not flat.  The probability that
a particle is found between $y$ and $y+dy$ is then $p(y)dy$, where
\begin{equation}
p(y) \, = \, \frac {\rho(y)} {N} \, = \, \int dy_c \, \pi(y_c) \,
q(2y_c-2y) \approx \pi(y),
\end{equation}
if $\pi(y)$ varies slowly compared to $q$ [$q(2y_c-2y)$ occurs because
the second particle of a pair has rapidity $y=2y_c-y$].  I then obtain
\begin{equation}
\rho^{(2)}(y_1,y_2) \, = \, [N(N-1)-N_c]p(y_1)p(y_2) + N_c \pi \left(
\frac {y_1+y_2} {2} \right) q(y_1-y_2),
\end{equation}
and consequently
\begin{equation}
R_2' (y;\Delta Y) \, \simeq \, 1 \, + \, \frac {n_c \, q(y)}
{dN/dy|_{y/2}},
\end{equation}
assuming that $dN/dy$ varies slowly so $\left[ dN/dy|_0 \right]
\left[ dN/dy|_y \right] \simeq \left[ dN/dy|_{y/2} \right]^2$.

The only difference from $R_2$ is that $dN/dy$ is now taken at
rapidity $y/2$.  The error will change slightly, as the error in
determining the correction factor should now be added to the
previous error.  The error in determining $p(y)$ is
\begin{equation}
e_p/p \, = \, 1/\sqrt{N_{ev} (dN/dy)_y \delta y},
\end{equation}
so the error in $R_2'$ is naively
\begin{equation}
e_R \, = \, \sqrt{6 / N_{ev} (dN/dy) \delta y},
\end{equation}
assuming that the distribution is approximately flat and that the
errors are uncorrelated.  This will lower the previous values of
$(s/n)_R$, and raise the numbers of events required for a good
measurement by about 50\%.

The most useful version of $S_2$ has only a simple additive correction
[\ref{rs2d}]:
\begin{equation}
S_2' (\delta y; \Delta Y) \, = \, S_2 (\delta y; \Delta Y) -
\Sigma_2 (\delta y; \Delta Y),
\end{equation}
where
\begin{equation}
\Sigma_2(\delta y; \Delta Y) = \frac {\displaystyle \Delta Y \,
\sum_{j=1}^{\Delta Y/\delta y} \,
\int_{(j-1) \delta y}^{(j-1/2) \delta y} dy_1 \, p(y_1) \,
\int_{(j-1/2) \delta y}^{j \delta y} dy_2 \, p(y_2)}
{\displaystyle \delta y \, \int_0^{\Delta Y/2} dy_1 \, p(y_1) \,
\int_{\Delta Y/2}^{\Delta Y} dy_2 \, p(y_2)} \, - \, 1.
\end{equation}

For the case where the particles are centered in the window,
$\int_0^{\Delta Y/2} dy \, p(y) \, = \, \int_{\Delta Y/2}^{\Delta Y}
dy \, p(y)$, the correction simplifies to
\begin{equation}
\Sigma_2(\delta y; \Delta Y) = \frac {4} {\Delta Y \delta y} \,
\sum_{j=1}^{\Delta Y/\delta y} \, \int_{(j-1) \delta y}^{(j-1/2) \delta y}
dy_1 \, \delta f(y_1) \, \int_{(j-1/2) \delta y}^{j \delta y} dy_2 \,
\delta f(y_2). \label{esig2c}
\end{equation}
Here $\delta f=p \Delta Y -1$ is the fractional difference from the mean
value.  As $\delta y \rightarrow 0$, $\Sigma_2 \rightarrow
\overline{\delta f^2}$ from below, so if the furthest bins are within
fraction $f$ of the mean density then $\Sigma_2 < f^2$.  The error
in the correction factor, $\Sigma_2$, is approximately the same size as
the error in $S_2$, so the experimental errors will be somewhat larger
than for the uncorrected case, just as for $R_2$.

\section{ Quantum interferometry } \label{sqi}

Experimenters are also very interested in using Hanbury/Brown-Twiss
interferometry to determine the geometry of the collision region in
high--multiplicity events [\ref{rhbt1}, \ref{rhbt2}].  This technique
uses species--dependent momentum correlation functions, for which
particle identification is required.  I briefly discuss statistical
aspects of the two most common methods for identical particle
interferometry, neglecting any difficulties due to errors in the
identification of particles or other technical problems.

Both techniques usually use pions that are identified in a spectrometer
covering $\Delta \Phi$ radians of azimuthal angle and $\Delta Y$ units
of pseudo--rapidity.  For pions, pseudo--rapidity and rapidity are
almost identical, so I do not differentiate between them in this paper.
The correlation function used is
\begin{equation}
C_2 (q) = \frac {\int_{0}^{\Delta \Phi} d\phi_1 d\phi_2
\int_{0}^{\Delta Y} dy_1 dy_2 \int dp_{T,1} dp_{T,2} \rho^{(2)}_{id}
\delta [q^2 + (p_1-p_2)^2]} {\int_{0}^{\Delta \Phi} d\phi_1 d\phi_2
\int_{0}^{\Delta Y} dy_1 dy_2 \int dp_{T,1} dp_{T,2}  \rho^{(2)}_{uc}
\delta [q^2 + (p_1-p_2)^2]},
\end{equation}
where $\delta$ is the Dirac $\delta$--function.  Here $\rho^{(2)}_{id}$
is the distribution for two identical particles, and $\rho^{(2)}_{uc}$
is an uncorrelated two--particle distribution; the two methods differ
only in the prescriptions used to construct $\rho^{(2)}_{uc}$.  I use
standard high energy units with $\hbar = c = 1$.

In the first method, referred to as event mixing, an uncorrelated
distribution is produced by combining particles (of the same species and
charge as those used in $\rho^{(2)}_{id}$) from different events.  This
is usually done by constructing simulated events using particles from
measured events, while ensuring that no two particles come from the
same events.  This process is very computationally intensive for
high--multiplicity events, as it is necessary to keep track of the event
from which each particle is taken and check all selections to ensure
that they don't come from an event that was used earlier.

Considerable computational difficulty is removed by constructing
$\rho^{(2)}_{uc}$ by convoluting the single--particle distribution
obtained by averaging over all events.  Using a sample of $N_{ev}$
events, each with $N$ particles in the spectrometer, this second
procedure yields
\begin{equation}
\rho^{(2)} (p_1, p_2) ~=~ \frac {N-1} {N} \, \rho (p_1) \, \rho (p_2)
{}~=~ \rho^{(2)}_{uc} \, + \, \frac {\rho^{(2)}_{id} \, - \,
\rho^{(2)}_{uc}} {N_{ev}},
\end{equation}
for all $p_1 \neq p_2$,
where $\rho^{(2)}_{uc}$ is the value obtained by taking all convolutions
with no two particles from the same event.  For $N_{ev} \gg 1$, the
difference between the proposed procedure and the usual one is small.
This change in procedure is even more important for measuring
higher--order correlations, as the construction of $\rho_{uc}$ by
the usual event mixing quickly becomes computationally prohibitive,
while the simple convolution of $\rho$ is almost always feasible.

In the second method, referred to as charge mixing, $\rho^{(2)}_{uc}$
is constructed using particles that are identical except for charge.
For example,
\begin{equation}
\rho^{(2)}_{uc} (p_1, p_2) ~=~ \rho^+ (p_1) \, \rho^- (p_2),
\end{equation}
where $\rho^{\pm}$ is the $\pi^{\pm}$ density, is often used for
pion interferometry.  As charge mixing does not require information from
more than one event, it is the preferred normalization technique for
analysis of single events.

Consider the analysis of a single event.  The mean number of $\pi^{\pm}$
mesons seen in a spectrometer covering $\Delta Y$ units of rapidity and
$\Delta \Phi$ radians of azimuthal angle is
\begin{equation}
N^{\pm} ~=~ \frac {(dN/dy) \, \Delta Y \, \Delta \Phi} {4 \pi},
\end{equation}
if particles are spread randomly in $y$ and $\phi$ with a uniform
distribution.  I imagine an ideal spectrometer that detects all
pions passing through it, independent of transverse momentum, $p_T$.
For simplicity, I assume a thermal--type $p_T$ distribution:
\begin{equation}
{\cal P} (p_T) ~=~ \frac {4 p_T} {\langle p_T \rangle^2} \,
e^{-2p_T/\langle p_T \rangle},
\end{equation}
where ${\cal P} (p_T) \, dp_T$ is the probability that a given pion has
transverse momentum between $p_T$ and $p_T+dp_T$, and $\langle p_T
\rangle$ is the mean $p_T$.

For a spectrometer large enough that edge effects are unimportant
($\Delta Y, \, \Delta \Phi \gg q/\langle p_T \rangle$), the standard
deviation of $C_2(q)$ is
\begin{equation}
\sigma_C ~=~ \frac {16 \pi} {(dN/dy) \, \Delta Y \, \Delta \Phi} \left\{
\frac {\langle I^2 \rangle} {\langle I \rangle^2} \, - \, 1 \,
+ \, \frac {\pi} {2 \, (dN/dy) \, q \, \delta q \, \langle I \rangle}
\right\},
\end{equation}
in the high--multiplicity limit, where
\begin{eqnarray}
\langle I^n \rangle = \int_0^{\infty} dp_{T,1} \, {\cal P} (p_{T,1}) \,
\Biggl[ \int_0^{\infty} dp_{T,2} \, {\cal P} (p_{T,2})
\int_{-\infty}^{\infty} dy \, \int_0^{2\pi} d\phi \nonumber \\
\delta \left( q^2 + 2 m^2 + 2 p_{T,1} p_{T,2} \cos \phi - 2
\sqrt {(p_{T,1}^2+m^2) (p_{T,2}^2+m^2)} \cosh y \right) \Biggr]^n.
\end{eqnarray}
Here $(\int_{q=0}^{q'} dq^2)^n \, \langle I^n \rangle / (\Delta Y \Delta
\Phi)^n$ is the probability that $n$ given particles each have momenta
within $q'$ of a particle at $y=\phi=0$, with transverse momentum
distribution ${\cal P} (p_{T,1})$.  The primary interest is in small $q$,
and $m^2 \ll \langle p_T \rangle^2$, so I evaluate $\langle I^n \rangle$
for $q^2 = m^2 =0$, obtaining
\begin{eqnarray}
\langle I \rangle ~\approx~ \frac {\left[ \Gamma (1/4) \right]^2 \,
\sqrt{\pi}} {2 \, m \, \langle p_T \rangle}, \\
\langle I^2 \rangle ~\approx~ \frac {\left[ \Gamma (1/4) \right]^4}
{m^2 \, \langle p_T \rangle^2},
\end{eqnarray}
and thus
\begin{equation}
\sigma_C ~=~ \frac {16 \pi} {(dN/dy) \, \Delta Y \, \Delta \Phi}
\left\{ \frac {4} {\pi} \, - \, 1 \, + \, \frac {\sqrt{\pi} \, m \,
\langle p_T \rangle} {[\Gamma (1/4)]^2 \, (dN/dy) \, q \, \delta q}
\right\}. \label{esigC2}
\end{equation}

The first term of eq.~(\ref{esigC2}) dominates as long as
\begin{equation}
q^2 \gg \frac {3 \sqrt{\pi} \, m \, \langle p_T \rangle} {\left[
\Gamma (1/4) \right]^2 \, (dN/dy)},
\end{equation}
where I have taken $\delta q = q$ for the smallest bin.  Using
$dN/dy=1000$ and $\langle p_T \rangle = 500$ MeV, I find that the
first term dominates as long as $q^2 \gg 30$ MeV$^2$, which is true
for all practical measurements.  Thus, the expected measurement error
is
\begin{equation}
e_C ~\approx~ \frac {4} {\sqrt {(dN/dy) \, \Delta Y \, \Delta \Phi}},
\label{eeC2} \end{equation}
independent of the momentum or the bin size!  As the signal is
approximately unity for $q=0$, a 4$\sigma$ determination is possible
with a single event for any spectrometer with $\Delta Y \Delta \Phi \,
> \, 256 \, (dN/dy)^{-1}$.

Quantum interference also produces two--particle rapidity correlations.
These correlations are due to collective effects, so they have a
different multiplicity dependence than the two--particle correlations
discussed in the previous section.  The momentum scale for quantum
interference $q^* \approx 1/r$, where $r$ is the size of the system, so
the rapidity scale is
\begin{equation}
y^* ~\approx~ \frac {q^*} {\langle p_T \rangle} ~\approx~ \frac {0.4 \,
{\rm fm}} {r},
\end{equation}
while the number of correlated pairs per particle is
\begin{equation}
k ~\approx~ \frac {{q^*}^2 \, \langle I \rangle \, (dN/dy)} {4 \pi}
{}~\approx~ \frac {0.5 \, (dN/dy) \, {\rm fm}^2} {r^2}.
\end{equation}
For simplicity, I assume that pairs are correlated if the momentum
difference is less than $q^*$, and uncorrelated otherwise.  Using
eqs.~(\ref{eS20}) and (\ref{egl0}), I obtain
\begin{equation}
S_2^{\rm be} (\delta y) ~\approx~ 1 \, + \, \frac {0.8 \, {\rm fm}^3}
{r^3 \, \delta y^2},
\end{equation}
for $\Delta Y \gg \delta y \gg 1 \, {\rm fm} / r$, while
eq.~(\ref{eglarge}) gives the maximum value,
\begin{equation}
S_2^{\rm be, max} ~\approx~ 1 \, + \, \frac {1.25 \, {\rm fm}} {r}.
\end{equation}

The two--particle correlation due to clusters is visible above
$S_2^{\rm be}$ as long as $dN/dy \, < \, 1.25 \, n_c \, y^* \, r^3 /
{\rm fm}^3$.  Single--event measurements of $S_2^{\rm be}$ are in principle
possible if $(dN/dy) \, \Delta Y \, > \, 40 \, r^2 / {\rm fm}^2$.  Recent
data [\ref{re802}] indicates that $r = 3 - 4$ fm for central Si collisions
at 14.6 GeV/nucleon (approximately the Si radius); using $r=7$ fm for U+U
collisions at RHIC and LHC, a single--event measurement of $S_2$ may be
possible.  However, such a measurement is very difficult technically,
requiring rapidity resolution to $y^* \approx 0.06$.

\section{ Higher--order correlation functions } \label{shocf}

It is also possible that higher--order correlation functions might give
better results (or more interesting results) than two--particle correlation
functions.  Higher--order correlation functions can in principle be
used to determine three--body and higher--order interactions, and many
experimenters have tried to use them for this purpose, although without
much success [\ref{rcum1}, \ref{rcum2}].  Alternatively, measuring
higher--order correlation functions might provide a more accurate
determination of the two--particle correlation function than can be
obtained from a direct measurement.

To test these hypotheses, I construct scaled factorial moments (SFMs)
[\ref{rsfms}], that also use the data more efficiently (although they
re--use pairs of particles).  For pedagogical purposes, I consider only
the so--called exclusive SFMs,
\begin{equation}
F_i(\delta y; \Delta Y) = \left( \frac {\Delta Y} {\delta y} \right)^{i-1}
\sum_{j=1}^{\Delta Y/\delta y}
\frac {\displaystyle \int_{(j-1) \delta y}^{j \delta y} dy_1 \, \cdots \,
dy_i \, \rho^{(i)} (y_1, \ldots , y_i)}
{\displaystyle N \cdots (N-i+1)}.
\end{equation}
Just as for $S_2$, $\Delta Y/\delta y$ must be an integer; however, any
two values of $\delta y$ use some common phase space, so it is impossible
to construct SFMs without over--using the available phase space.

To examine the feasibility of studying higher--order correlations,
I extend my previous approximations to include three--body correlations,
assuming $N_t$ correlated triplets with the boost--invariant distribution
$q_t (y_1-y_2, y_1-y_3)/\Delta Y$.  For $\Delta Y \gg y^*$, the maximum
value of $F_i$ is
\begin{equation}
F_i^{\rm max} ~\approx~ 1 \, + \, \left[ \frac {i(i-1)} {2} \right]
\frac {k} {(dN/dy)y^*} \, + \, \left[ \frac {i(i-1)(i-2)} {6}
\right] \frac {k_t} {\left[ (dN/dy) y^* \right]^2},
\end{equation}
where $k_t = N_t/N$.  For cluster decay, $k_t \approx n_c^2$ for $n_c
\gg 1$, so
\begin{equation}
\left[ \frac {(i(i-1)(i-2)} {6} \right] \frac {k_t} {\left[ (dN/dy) y^*
\right]^2} ~\approx~ \frac {2(i-2)} {3i(i-1)} \left[ F_i-1 \right]^2,
\label{etrip} \end{equation}
is the three--particle contribution to $F_i$, while $F_i-1$ is the
two--particle contribution.

The observed three--particle correlation decreases with increasing
multiplicity as $(dN/dy)^{-2}$, even faster than the two--particle
correlation.  Equation~(\ref{etrip}) is apparently very general, as it
holds for any values of $n_c$ and $y^*$.  Given a distribution of values,
corrections of order unity are likely; however, it is still probable that
the difficulty of extracting the three--particle correlation increases
very fast as the two--particle correlation function decreases.  This is,
in my opinion, the most likely reason for the failure of experimenters to
extract significant three--particle correlations [\ref{rcum1},
\ref{rcum2}] from their data, as virtually all two--particle correlations
are approximately 10\% or smaller.

The second hypothesis, that higher--order correlation functions might
provide a more accurate determination of the two--particle correlations
than direct measurement, seems to be true.  Under the assumptions that
I use to calculate $R_2$,
\begin{equation}
F_i (\delta y) \, = \, 1 \, + \, \left[ \frac {i(i-1)} {2} \right] \,
\frac {k} {dN/dy} \, \left[ h(\delta y) - h(\Delta Y) \right],
\end{equation}
where
\begin{equation}
h(z) \, = \, \frac {2} {z^2} \, \int_0^z dx \, (z-x) \, q(x) \,
\approx \, q(z/3).
\end{equation}
I calculate the error in the same manner as before, assuming that there
are no correlations:
\begin{equation}
\sigma_{F_i} \, = \, \frac {i^2} {N_{ev} \, N} \, \left[ 1 \,
+ \, {\cal O} \left( \frac {\delta y} {\Delta Y} \right) \right].
\end{equation}

The statistical error in $F_i$ is proportional to $i$, while the signal is
proportional to $i(i-1)$, so $s/n$ is proportional to $i-1$ and thus
improves with increasing $i$.  However,
the above arguments are valid only when $(dN/dy) \, \delta y \gg i$,
in which case most bins contain $i$ particles, and even then
apply only to statistical noise.  In most cases, there is also noise
from undesired correlations produced by the detector, and this systematic
noise is proportional to $i(i-1)/2$.  It is thus possible that $s/n$ is
approximately independent of $i$.

It appears that measuring higher--order correlation functions is the
most accurate way to determine two--particle correlations.  Because
SFMs re--use data, the cleanest approach is probably to use higher--order
SBCFs [\ref{rsbcfs}]:
\begin{equation}
S_i(\delta y; \Delta Y) = \frac {\displaystyle \Delta Y^{i-1} \,
\sum_{j=1}^{\Delta Y/\delta y} \, \prod_{n=1}^i \,
\int_{[j-n/i] \delta y}^{[j-(n-1)/i] \delta y} dy_n \,
\rho^{(i)} (y_1, \ldots, y_i)}
{\displaystyle \delta y^{i-1} \, \prod_{n=1}^i \,
\int_{(n-1) \Delta Y / i}^{n \Delta Y / i} dy_n \,
\rho^{(i)} (y_1, \ldots, y_i)}.
\end{equation}
Maximal use of the data without re--use is simple for $S_2$, but not
easily achieved for high--order SBCFs.  As a result, it may be preferable
to use some different generalization of $S_2$ for maximal efficiency.
Higher--order correlation functions might also give better results for
quantum interference measurements than the commonly used two--particle
correlation functions.

\section{ Conclusions } \label{sc}

I have discussed four types of correlation measurements: hot spot
searches, two--particle rapidity correlations, two--particle momentum
correlations (for quantum interferometry), and higher--order correlation
functions.  Hot spot searches are most likely to be profitable in events
of high energy but relatively low multiplicity.  Two--particle rapidity
correlations are most easily measured in events of relatively low
multiplicity, if high--multiplicity events are just superpositions of
lower--multiplicity events.  A good measurement of $S_2$ at RHIC or LHC
will require at least 800 events.

Single--event measurement of two--particle momentum correlations due to
quantum interference is possible in high--multiplicity events
with spectrometer coverage $\Delta Y \Delta \Phi > 256 \, (dN/dy)^{-1}$,
which should be easily attainable at RHIC and LHC.  Rapidity correlations
due to quantum interference are in principle measurable in single events
at RHIC and LHC, but such measurements would be very difficult technically.
Measuring higher--order correlation functions in high--multiplicity events
gives little information about three--body and higher--order correlations.
However, measuring higher--order correlation functions can give a more
accurate determination of two--particle correlations than direct
measurement of two--particle correlation functions.

\ulsect{Acknowledgements}

I thank J. Kapusta and G. Fai for critical reading of this manuscript,
and the Theoretical Physics Institute of the University of Minnesota
for their hospitality during part of the time I worked on this paper.
This work was supported in part by the US Department of Energy under
Grants No.\ DOE/DE-FG02-86ER-40251 and DOE/DE-FG02-87ER-40328.

\ulsect{References}

\begin{list}{\arabic{enumi}.\hfill}{\setlength{\topsep}{0pt}
\setlength{\partopsep}{0pt} \setlength{\itemsep}{0pt}
\setlength{\parsep}{0pt} \setlength{\leftmargin}{\labelwidth}
\setlength{\rightmargin}{0pt} \setlength{\listparindent}{0pt}
\setlength{\itemindent}{0pt} \setlength{\labelsep}{0pt}
\usecounter{enumi}}

\item T.J. Humanic, for the NA35 Collaboration (A. Bamberger {\it et al.}),
Z. Phys.\ C {\bf 38}, 79 (1988). \label{rsea}

\item R. Stock, Nucl.\ Phys.\ {\bf A498}, 333c (1989).

\item J. Harris, in {\it Advances in Nuclear Dynamics}, Proceedings of
the 8$^{\rm th}$ Winter Workshop on Nuclear Dynamics, Jackson Hole,
Wyoming, January 1992 (World Scientific, 1992: W. Bauer and B. Back,
eds.). \label{rseo}

\item K.L. Kowalski and C.C. Taylor, Case Western Reserve University
preprint CWRUTH--92--6 (June 1992). \label{rkt}

\item D. Seibert, Kent State University preprint KSUCNR--009--92 (July
1992), to be published in the proceedings of the VIIth International
Symposium on Very High Energy Cosmic Ray Interactions, University of
Michigan, Ann Arbor, Michigan, June 1992 (L.W. Jones, ed.) \label{raa}

\item For a good review, see L. Fo\`a, Phys.\ Rep.\ {\bf 22}, 1 (1975).
\label{rr2}

\item D. Seibert, Phys.\ Rev.\ D {\bf 41}, 3381 (1990). \label{rmult}

\item KLM Collaboration, R. Holynski {\it et al.}, Phys.\ Rev.\ Lett.\
{\bf 62}, 733 (1989). \label{rklm}

\item EMU01 Collaboration, E. Stenlund {\it et al.}, Nucl.\ Phys.\
{\bf A498}, 541c (1989). \label{remu01}

\item D. Seibert, Phys.\ Rev.\ Lett.\ {\bf 63}, 136 (1989). \label{rcl1}

\item S. Voloshin and D. Seibert, Phys.\ Lett.\ B {\bf 249}, 321 (1990);
D. Seibert and S. Voloshin, Phys.\ Rev.\ D {\bf 43}, 119 (1991).
\label{rsbcfs}

\item D. Seibert, Phys.\ Rev.\ C {\bf 44}, 1223 (1991). \label{rs2d}

\item P.V. Ruuskanen and D. Seibert, Phys.\ Lett.\ B {\bf 213}, 227
(1988).\label{rRS}

\item S. Pratt, Phys.\ Rev.\ D {\bf 33}, 1314 (1986). \label{rhbt1}

\item W.G. Gong {\it et al.}, Phys.\ Rev.\ Lett.\ {\bf 65}, 2114
(1990). \label{rhbt2}

\item E--802 Collaboration, T. Abbott {\it et al.}, Phys.\ Rev.\ Lett.\
{\bf 69}, 1030 (1992). \label{re802}

\item P. Carruthers, H.C. Eggers, and I. Sarcevic, Phys.\ Rev.\ C
{\bf 44}, 1629 (1991). \label{rcum1}

\item H.C. Eggers, Ph.D. thesis, University of Arizona, 1991
(unpublished).\label{rcum2}

\item A. Bia\l as and R. Peschanski, Nucl.\ Phys.\ {\bf B273}, 703
(1986). \label{rsfms}

\end{list}

\vfill \eject

\end{document}